\documentclass[amssymb,jap,twocolumn,superscriptaddress,floats,showkeys,showpacs]{revtex4-1}

\usepackage{float}
\usepackage[caption=false]{subfig}
\usepackage{ulem}
\usepackage[euler]{textgreek}
\usepackage{mathrsfs}
\usepackage[T1]{fontenc}
\usepackage{bm}
\usepackage{graphicx}
\usepackage{natbib}
\usepackage{amsmath}
\usepackage{textcomp}
\usepackage{epstopdf}
\usepackage[dvipsnames]{xcolor}
\definecolor{MMgreen}{RGB}{0,128,0}
\usepackage{hyperref} 
\hypersetup{colorlinks,citecolor=blue, filecolor=blue, linkcolor=blue , urlcolor=blue}
\usepackage{sidecap}
\newcommand{\unit}[1]{\ensuremath{\, \mathrm{#1}}}

\def \FUW{Institute of Experimental Physics, Faculty of Physics, University of Warsaw, ul. Pasteura 5, 02-093 Warsaw, Poland}
\def \Grenoble{Laboratoire National des Champs Magn\'etiques Intenses, CNRS-UGA-UPS-INSA-EMFL, 25, avenue des Martyrs, 38042 Grenoble, France} 
\def \Prague{Department of Condensed Matter Physics, Faculty of Mathematics and Physics, Charles University in Prague, Ke Karlovu 5, Praha 2 CZ-121 16, Czech Republic}
\def \Brno{Central European Institute of Technology, Brno University of Technology,  Purky\v{n}ova 656/123, 612 00 Brno, Czech Republic}
\def \Wroclaw{Department of Experimental Physics, Faculty of Fundamental Problems of Technology, Wroc\l{}aw University of Science and Technology, ul. Wybrze\.ze Wyspia\'nskiego 27, 50-370 Wroc\l{}aw, Poland}
\def \Kenji{Research Center for Functional Materials, National Institute for Materials Science, 1-1 Namiki, Tsukuba 305-0044, Japan}
\def \Takashi{International Center for Materials Nanoarchitectonics, National Institute for Materials Science, 1-1 Namiki, Tsukuba 305-0044, Japan}

\begin{document}

\title{Valley polarization of singlet and triplet trions in WS$_2$ monolayer in magnetic fields}

\author{P. Kapu\'sci\'nski}\email{piotr.kapuscinski@lncmi.cnrs.fr}\affiliation{\Grenoble}\affiliation{\Wroclaw}
\author{D. Vaclavkova}\affiliation{\Grenoble}
\author{M. Grzeszczyk}\affiliation{\FUW}
\author{A.~O.~Slobodeniuk}\affiliation{\Grenoble}\affiliation{\Prague}
\author{K. Nogajewski}\affiliation{\FUW}
\author{M. Bartos}\affiliation{\Grenoble}\affiliation{\Brno}
\author{K.~Watanabe}\affiliation{\Kenji}
\author{T. Taniguchi}\affiliation{\Takashi}
\author{C. Faugeras}\affiliation{\Grenoble}
\author{A. Babi\'nski}\affiliation{\FUW}
\author{M. Potemski}\affiliation{\Grenoble}\affiliation{\FUW}
\author{M. R. Molas}\email{maciej.molas@fuw.edu.pl}\affiliation{\FUW}

\begin{abstract}
	
The spectral signatures associated with different negatively charged exciton complexes (trions) in a WS$_2$ monolayer encapsulated in hBN, are analyzed from low temperature and polarization resolved reflectance contrast (RC) and photoluminescence (PL) experiments, with an applied magnetic field. Based on results obtained from the RC experiment, we show that the valley Zeeman effect affects the optical response of both the singlet and the triplet trion species through the evolution of their energy and of their relative intensity, when applying an external magnetic field. Our analysis allows us to estimate a free electron concentration of $\sim 1.3 \cdot 10^{11}$~cm$^{-2}$. The observed evolutions based on PL experiments on the same sample are different and can hardly be understood within the same simple frame highlighting the complexity of relaxation processes involved in the PL response.  
	
\end{abstract}

\maketitle

\section{Introduction \label{sec:Intro}}

The charged excitons or trions as consisting of an electron-hole pair and an excess carrier (an electron or a hole) are the simplest complexes to investigate many-body phenomena. In conventional 2D semiconducting structures, such as GaAs and CdTe quantum wells, it is known that at zero magnetic field only a singlet state of charged exciton can be bound due to the Pauli exclusion principle. Nonetheless, the observation of a triplet state of trion is possible under an applied magnetic field~\cite{shield,hadas,astakhov}. For so-called darkish monolayers (MLs) of semiconducting transition metal dichalcogenides (S-TMDs), $i.e.$ WSe$_2$ and WS$_2$, possessing an optically inactive (dark) ground exciton state~\cite{molas,Zhou2017,Zhang2017,wang2017}, both the singlet and triplet states of the negatively charged excitons can be observed at zero magnetic field~\cite{boulesbaa,plechingerTRION,molasNanoscale,jadczaknano}. This is possible due to the fact that, in these materials, the direct band gap is located at the two nonequivalent K$^{+}$ and K$^{-}$ valleys, what allows for the formation of triplet trions involving two electrons with the same spin, but originating from opposite valleys. Moreover, thanks to the strong Coulomb interactions in these materials, triplet and singlet states of trions can be energetically resolved, with energy splitting of few meV ~\cite{jakubczyk,vaclavkova,jadczak2019}. 

In this work, we investigate the magnetic field evolution of the singlet and triplet states of the negatively charged excitons in WS$_2$ monolayer encapsulated in hexagonal BN (hBN) by means of reflectance contrast (RC) and photoluminescence (PL) experiments. We demonstrate that the absorption of both the singlet and triplet trions, probed by the RC experiment, becomes strongly polarized when magnetic field is applied. 
We describe this effect as a consequence of the magnetic field induced valley polarization of the free carriers in the monolayer. The results of PL experiments on the same structure show different results.
It reveals that PL measurements do not provide a reliable information about valley-polarization of free carriers. Furthermore, we show that g-factors of both trion complexes have the different values when probed by the RC and the PL experiments, suggesting their sensitivity to photodoping.

\section{Sample and experimental setup \label{experiment}}

The studied sample is composed of a WS$_2$ monolayer encapsulated in hBN and deposited on a bare Si substrate. The structure was produced by a two-stage polydimethylsiloxane (PDMS)-based~\cite{gomez} mechanical exfoliation of WS$_2$ and hBN bulk crystals. The bottom hBN layer is first obtained by non-deterministic exfoliation on the Si substrate. The structure is then realized by successive deterministic dry transfers of the WS$_2$ ML followed by the capping hBN flake onto the bottom hBN flake. 

The $\mu$-PL and $\mu$-RC measurements were performed using a CW diode laser ($\lambda$=515~nm) and a 100~W tungsten halogen lamp, respectively. Magneto-optical experiments were performed in the Faraday configuration using an optical-fiber-based insert placed in a superconducting coil producing magnetic fields up to 14 T. The sample was mounted on top of a $x$-$y$-$z$ piezo-stage kept in gaseous helium at $T$=4.2 K. The excitation light was coupled to an optical fiber with a core of 50 $\mu$m diameter and focused on the sample by an aspheric lens (spot diameter around 10 $\mu$m). The signal was collected by the same lens, injected into a second optical fiber of the same diameter, and analyzed by a \mbox{0.5 m} long monochromator equipped with a CCD camera. A combination of a quarter wave plate and a polarizer were used to analyse the circular polarization of signals. The measurements were performed with a fixed circular polarization, whereas reversing the direction of magnetic field yields the information corresponding to the other polarization component due to time-reversal symmetry.

We define the RC spectrum as $RC(E)=\frac{R(E)-R_0(E)}{R(E)+R_0(E)}\times 100\%$, where $R(E)$ and $R_0(E)$ are the reflectance of the sample and of the same structure without the WS$_2$ monolayer, respectively.

\section{Experimental results \label{results}}
\begin{figure}[t]
	\includegraphics[width=\columnwidth]{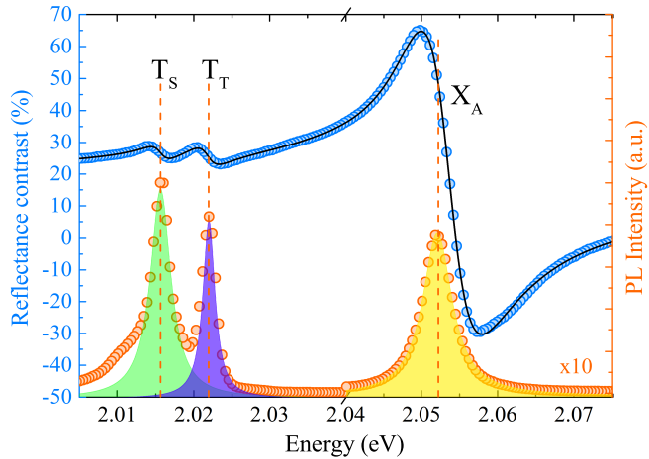}
	\caption{Reflectance contrast (blue dots) and photoluminescence (orange dots, $E_{exc}=2.408 \: \unit{eV}$, $P_{exc}=20 \:  \unit{\mu W}$) spectra of WS$_2$ monolayer encapsulated in hBN, measured at $T=4.2 \: \unit{K}$. The black curve is the fitting result of the RC spectrum using three Fano resonances. The purple/green and yellow Lorentzians represent the best fits to emission lines due to the singlet/triplet trions and neutral exciton, respectively. The PL spectrum in the region of X$\unit{_A}$ peak was multiplied by 10 for clarity. }
	\label{fig:fig1}
\end{figure}
\subsection{Trion resonances in monolayer WS$_2$ \label{results:resonances}}
The measured low-temperature ($T=4.2 \unit{K}$) reflectance contrast (RC) and photoluminescence (PL) response of the WS2 monolayer encapsulated in hBN flakes, in the range of energy of exciton and trion complexes, are presented in Fig.~\ref{fig:fig1}. The RC spectrum is composed of three distinct resonances, denoted as X$\unit{_A}$, T$\unit{_T}$, and T$\unit{_S}$. These transitions are also visible in the emission spectrum. The X$\unit{_A}$ resonance can be assigned to the neutral exciton in agreement with other studies of WS$_2$ monolayers ~\cite{chernikov,scrace,chernikov2015,molasNanoscale,jadczaknano,Molas5s}. Following the previous observations~\cite{plechingerTRION, molasNanoscale, vaclavkova,jadczak2019}, the T$\unit{_T}$ and T$\unit{_S}$ features are identified as the triplet and singlet states of the negatively charged excitons (trions), respectively.

\begin{figure}[!b]
	\includegraphics[width=\columnwidth]{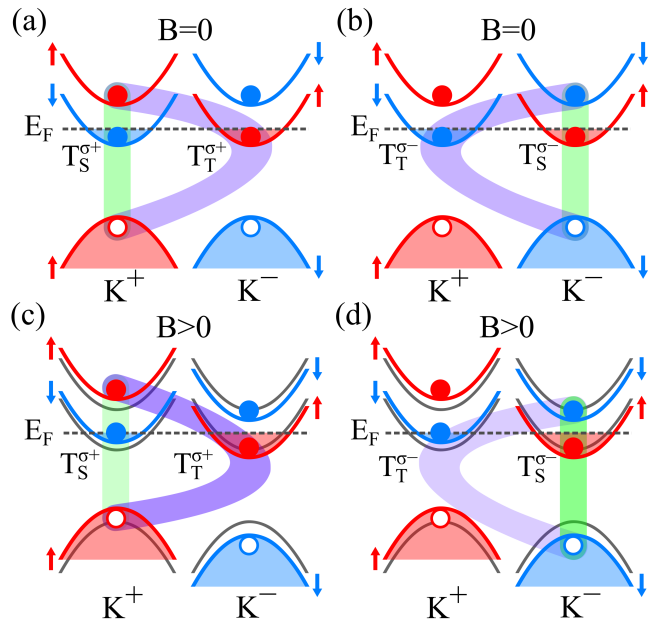}
	\caption{Schematic illustration of the circularly-polarized (a)/(c) $\sigma^+$ and (b)/(d) $\sigma^-$ singlet (T$\unit{_S}$) and triplet (T$\unit{_T}$) trions formed at K$^{+}$ and K$^{-}$ points of the Brillouin zone at 	zero/non-zero magnetic field.  The red (blue) curves 	indicate the spin-up (spin-down) subbands. The grey curves in (c) and (d) denote the position of subbands at zero field. The electrons (holes) in the conduction (valence) band are represented by closed (open) circles. Saturation of green and violet colours is related to the probability of trions formation.}
	\label{fig:figX}
\end{figure}

A schematic representation of the spin and valley configurations of these two distinct trion complexes at zero magnetic field is shown in Fig.~\ref{fig:figX}(a) and (b). To extract the energies and relative amplitudes of these transitions, the RC and PL spectra were fitted using three Fano-type and Lorentzian functions, respectively. The T$\unit{_T}$ absorption/emission resonances are located $\sim$30 meV below the X$\unit{_A}$, while the energy separation $\Delta\unit{_{T-S}}$, between T$\unit{_T}$ and T$\unit{_S}$ is of $\sim$7 meV. These values are in a very good agreement with those previously reported in Refs ~\cite{jakubczyk,vaclavkova,jadczak2019, AroraPRL2019, arora2019dark}. Note that the   PL spectrum also includes several additional lines (not shown here) apparent on the low energy side of trions emissions, characteristic of the recombination of various excitonic complexes, such as dark excitons, biexcitons, charged biexcitons and localized excitons ~\cite{molas,plechinger,jadczaknano,molasNanoscale,Paur2019}.

The observed differences in terms of relative amplitude, of the trions and exciton responses in the RC and in the PL spectra of the structure highlights the fundamental difference in these two optical processes: RC is a probe of the density of states and of the oscillator strength of a given transition, while the PL intensity also reflects occupation of the states and hence strongly depends on the efficiency of relaxation processes. While the neutral exciton formation is related to the density of excitonic states and to the intrinsic oscillator strength characterising a given material ~\cite{li, arora2015, aroramose2}, an additional free electron is required to create negative trions. The observation of trion related features in the optical response hence strongly depends on the carrier concentration or on the Fermi energy ($E_F$) which can be tuned externally for instance in gated structures ~\cite{chernikov2015, shang, molasNanoscale, Paur2019}. 
In our structure, there is most probably an unintentional n-type doping, commonly reported in WS$_2$ ~\cite{plechinger, molasNanoscale, jadczaknano, plechingerTRION}, allowing for the formation of trions. As the amplitude of the trions resonances in the RC response is rather small compared to exciton resonance, the free carrier concentration is expected to be relatively small.

\subsection{Magnetic field effect on the trions resonances  \label{results:magnet}}

\begin{figure*}[t]
	\includegraphics[width=\textwidth]{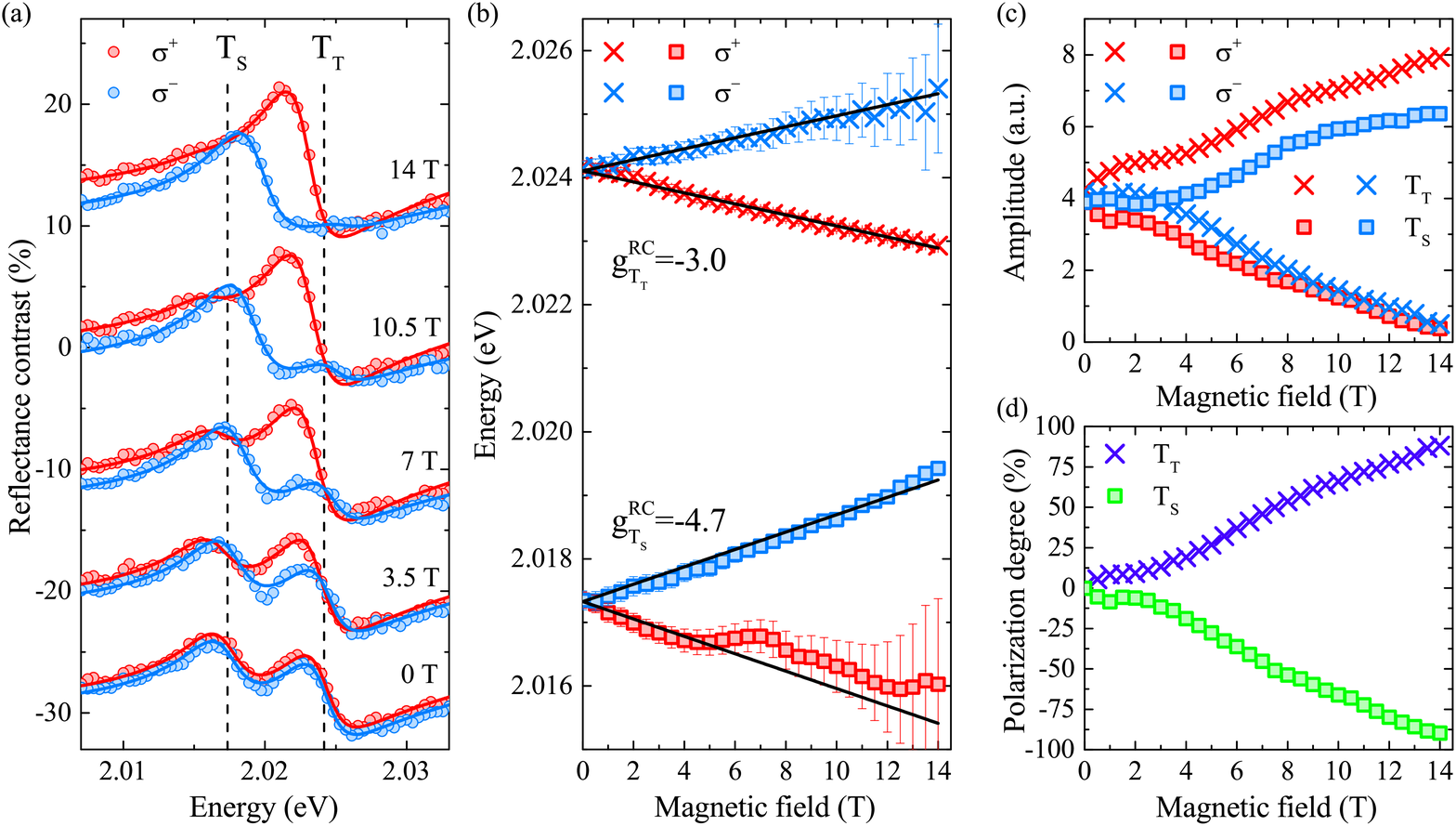}
	\caption{(a) Helicity-resolved low temperature ($T=4.2 \: \unit{K}$) reflectance contrast spectra of singlet (T$\unit{_S}$) and triplet (T$\unit{_T}$) trions measured on the encapsulated WS$_2$ monolayer at selected magnetic fields. The dots represent measured signals, while the lines correspond to fitted curves. The red (blue) colour correspond to the $\unit{\sigma^{+}}$-polarized ($\unit{\sigma^{-}}$-polarized) spectra. The dashed vertical lines show zero-field energies of trions. The spectra are vertically shifted factors for clarity. 
	(b) Transition energies of the T$\unit{_S}$ and T$\unit{_T}$ features extracted from the fits to the RC spectra as a function of magnetic field. The squares (crosses) correspond to the T$\unit{_S}$ (T$\unit{_T}$). The solid black curves represent fits with Eq.~\ref{eq:1}. Magnetic field evolutions of (c) the intensities and (f) the polarization degrees of the T$\unit{_S}$ and T$\unit{_T}$ transitions extracted from the fits to the RC spectra.
	}
	\label{fig:fig3}
\end{figure*}

Fig.~\ref{fig:fig3} presents the magnetic field evolution of the helicity resolved RC response in the range of energy of both T$\unit{_T}$ and T$\unit{_S}$ resonances, measured on the encapsulated  WS$_2$ ML up to $B = 14$ T. Applying a magnetic field yields two main effects visible of these RC traces: a splitting of both T$\unit{_T}$ and of T$\unit{_S}$ resonances into two circularly polarized components, and a strong variation of their intensities. The splitting is a consequence of the excitonic Zeeman effect, while the description of the intensities and of their evolution is more complex: The intensity of T$\unit{_T^{\sigma -}}$ and of T$\unit{_S^{\sigma +}}$ completely vanishes in the RC experiment while T$\unit{_T^{\sigma +}}$ and T$\unit{_S^{\sigma -}}$ increase up to $B = 14$ T. To analyse in details these evolutions, we have fitted these results with Fano-like functions. The obtained evolutions of both the energies and of the intensities of $\unit{\sigma^{+}}$ and $\unit{\sigma^{-}}$ polarized T$\unit{_S}$ and T$\unit{_T}$ components in magnetic fields are shown in Figs ~\ref{fig:fig3} (b) and (c). 

\begin{figure*}[t]
	\includegraphics[width=\textwidth]{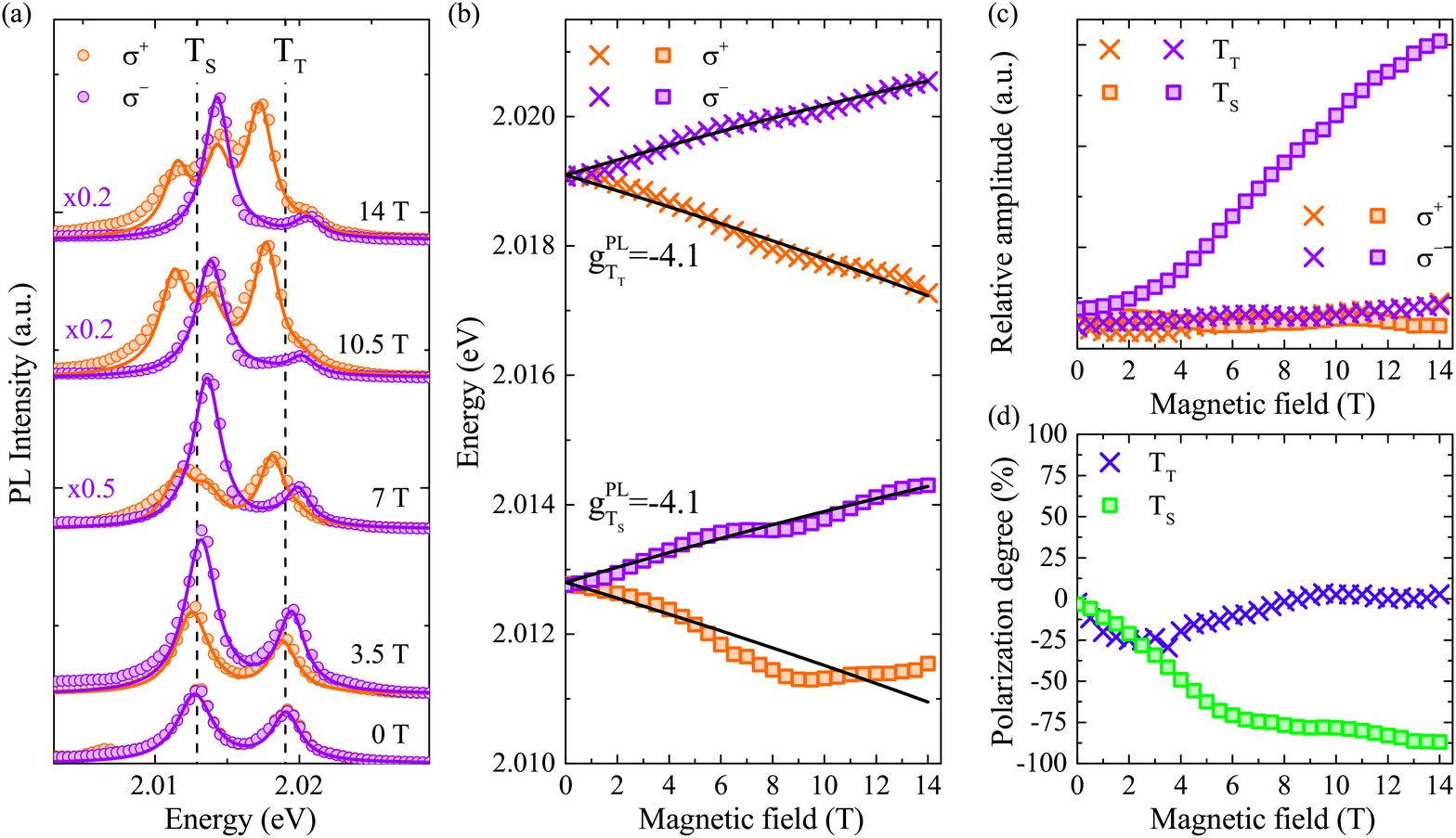}
	\caption{(a) Helicity-resolved low temperature ($T=4.2 \: \unit{K}$) photoluminescence ($E_{exc}=2.408 \: \unit{eV}$, $P_{exc}=20 \:  \unit{\mu W}$) spectra of singlet (T$\unit{_S}$) and triplet (T$\unit{_T}$) trions measured on the encapsulated WS$_2$ monolayer at selected magnetic fields. The dots represent measured signals, while the lines correspond to fitted curves. The orange (purple) colour correspond to the $\unit{\sigma^{+}}$-polarized ($\unit{\sigma^{-}}$-polarized) spectra. The dashed vertical lines show zero-field energies of trions. The spectra are vertically shifted factors for clarity and some detected at $\sigma^-$ were multiplied by indicated scaling factors for clarity. 
	(b) Transition energies of the T$\unit{_S}$ and T$\unit{_T}$ features extracted from the fits to the PL spectra as a function of magnetic field. The squares (crosses) correspond to the T$\unit{_S}$ (T$\unit{_T}$). The solid black curves represent fits with Eq.~\ref{eq:1}. Magnetic field evolutions of (c) the intensities and (f) the polarization degrees of the T$\unit{_S}$ and T$\unit{_T}$ transitions extracted from the fits to the PL spectra.
	}
	\label{fig:fig4}
\end{figure*}

The T$\unit{_T}$ and T$\unit{_S}$ lines split into two circularly-polarized components, where $\unit{\sigma^{+}}$/$\unit{\sigma^{-}}$ features experiences redshift/blueshift with increasing magnetic field, as it is presented schematically in Fig.~\ref{fig:figX}(c) and (d). For a comprehensive description of the effect of an external magnetic field on the energies of the conduction and valence bands see Ref. ~\cite{koperskiMagneto}.
The valley Zeeman effect for trions can be described by the following equation:

\begin{equation} \label{eq:1}
E(B)=E_{0} \pm \dfrac{1}{2}g \mu_{B} B.
\end{equation}

where, $E_{0}$ is the energy of the transition at zero field, $g$ denotes the g-factor of the considered excitonic complex and $\mu_{B}$ is the Bohr magneton. The results of fitting are shown in Fig.~\ref{fig:fig3} (b) as solid black curves. We find that the g-factors for the singlet and triplet trions differ significantly with $g\unit{^{RC}_{T_S}} \approx -4.7$ and $g\unit{^{RC}_{T_T}} \approx -3.0$. The corresponding g factor of the X$\unit{_A}$ transition is of the order of $g\unit{^{RC}_{X_A}} \approx -4.2$ from our RC spectra. 

Let us first discuss the evolution of the intensities of the different trion complexes components. Each band has an associated magnetic moment build up from three contributions namely the spin contribution, the Berry curvature contribution and the atomic orbital contribution. When a magnetic field is applied depending on the K$^{+}$/K$^{-}$ valley in which they reside, electronic bands in both the conduction and the valence band will shift in energy, giving rise to the valley Zeeman effect. Bands in K$^{+}$ valley will increase in energy, while bands in K$^{-}$ valley will decrease in energy.

The evolutions of the intensities of the different trion components find their explanation in the energy shifts of the different bands induced by the magnetic field relative to the Fermi energy, considered as constant in our uncontacted structure. In Fig.~\ref{fig:fig3} (c), the intensities of T$\unit{_T^{\sigma +}}$ and of T$\unit{_S^{\sigma -}}$, deduced from RC experiments, increase significantly when increasing the magnetic field. These two components imply an additional electron in the K$^{-}$ valley (see Fig.~\ref{fig:figX}(a) and (b)). Simultaneously, the intensities of the two other T$\unit{_T^{\sigma -}}$ and  T$\unit{_S^{\sigma +}}$ trion complexes vanish. If we consider a weak residual doping of the monolayer with a Fermi energy slightly above the bottom conduction band, and below the second conduction band (see Fig.~\ref{fig:figX}(c) and (d)), then a magnetic field can have a profound effect on the free carrier concentration in the two valleys. With a fixed Fermi energy, the free electron concentration in K$^{+}$ valley decreases gradually when applying a magnetic field, as the bottom of the conduction band rises towards the Fermi energy because of the valley Zeeman effect. The probability for the creation of a trion with an additional electron in the K$^{+}$ valley, directly related to the free carrier concentration, strongly decreases and should finally vanish when the bottom conduction band is tuned above the Fermi energy.

In order to investigate the details of the effect of a magnetic field on the measured amplitudes of trion complexes, we have calculated the degree of circular polarization defined as $\mathscr{P}=(I_{\sigma^+}-I_{\sigma^-})/(I_{\sigma^+}+I_{\sigma^-})\times 100\%$, where $I_{\sigma^+}$/$I_{\sigma^-}$ is the signal intensity detected in the $\sigma^+$/$\sigma^-$ polarization. Determined from  the RC experiments, the degrees of circular polarization of T$\unit{_T}$ and T$\unit{_S}$ are opposite, and they grow almost linearly when increasing the magnetic field, with a rate close to $\sim 6.6 \%.$T$^{-1}$, reaching values as high as 90$\%$ at $B=14$ T (see Fig.~\ref{fig:fig3}(d)). In a first approximation, the density of states of the conduction band of this two-dimensional electronic system is constant as a function of energy and, because the energy shift of the band is linear in magnetic field, the polarization of T$\unit{_T}$ and T$\unit{_S}$ is also expected to grow linearly with the magnetic field and then, to saturate at 100$\%$ when the bottom conduction band in the K$^{+}$ valley reaches the Fermi energy (not seen in our experiment). As the energy of the bottom conduction band  at the K$^{+}$ valley of WS$_2$ grows as $\mu_B B$, the saturation of $\mathcal{P}$ is expected when $\mu_B B_c = E_F$, where $B_c \sim 15.2$ T allows to estimate in our case $E_F \sim 0.9$~meV. Now, using the relation $n=m_e E_F/{\pi \hbar^2}$ and assuming the effective mass of electrons $m_e \sim 0.35$ (taken from Ref. \cite{kormanyos}), we can estimate the free carriers concentration to be of the order of $n \sim 1.3 \cdot 10^{11} \unit{cm^{-2}}$. This result shows that free electrons in the conduction band can be valley polarized with an external magnetic field, and that this effect determines the degree of circular polarization of the different trion species as probed by RC spectroscopy.

In other words, the Fermi energy of the system is fixed, but the carrier densities in each valley is changing when a magnetic field is applied, free carriers from the K$^{+}$ valley being transferred to the K$^{-}$ valley because of the valley Zeeman effect. We argue that this effect can also explain the different g-factors extracted for the singlet and triplet trion species. Indeed, optical investigations of high quality gated monolayers have shown that the trion energies depends on the free carrier density \cite{SmolenskiMoSe2} and that, if the Fermi energy lies in the conduction band, the trio  energy increases when the free carrier density is increased. When a magnetic field is applied, the bands in the valleys move in the way indicated in Fig.~\ref{fig:figX}(c) and (d), and, as explained in the preceding paragraphs, free carriers are gradually transferred from K$^{+}$ to K$^{-}$ valley. This implies that, in addition to the valley Zeeman effect, we are also facing a change in the trion complexes energy induced by the charge redistribution in the two valleys. The free carrier density in the K$^{+}$ valley decreases under magnetic field, which leads to a decrease of energy the two trions based on this particular population of free carriers, namely the T$\unit{_T^{\sigma -}}$ and the T$\unit{_S^{\sigma +}}$. Simultanuously, the concentration of free carriers in the K$^{-}$ valley increases by a similar amount, leading to an increase of the energy of both T$\unit{_T^{\sigma +}}$ and the T$\unit{_S^{\sigma -}}$ components. This effect leads to an apparent increase in the g-factor of the singlet trion and to an apparent decrease of the triplet trion, as evidenced in the evolutions presented in Fig.~\ref{fig:fig3}(b).

To conclude this discussion of the results of magneto-RC experiments, both the evolutions of the intensities (or of the degree of circular polarization) and of the g-factors of the two trion species can be understood considering magnetic field induced free carrier redistribution in the two valleys of our ungated WS$_2$ monolayer. 

\subsection{Discussion of the PL response}
We now turn to the photoluminescence investigation of the trion complexes. Typical polarization resolved PL spectra at selected values of the magnetic field are presented in Fig.~\ref{fig:fig4}(a). We have used lorentzian functions to fit the evolution of their amplitude and energy as a function of the magnetic field. Similar to RC experiments, we define a degree of circular polarisation based on the amplitude of the PL signals, but relative to that of A-exciton to remove the Faraday effect naturally affecting the experimental data. The results of this analysis are displayed in Fig.~\ref{fig:fig4}(b), (c) and (d) and they differ significantly from the previously discussed case of RC experiments. The first difference lies in the values of the g factor for both singlet and triplet trions. The PL investigation indicates that both trions have identical g-factors of $\sim -4.1$, close to that of the neutral A-exciton. The second difference is in the evolution of the intensity of the different polarization resolved components of the trions. Only the T$\unit{_S^{\sigma -}}$ component increases significantly, by a factor 10 up to $B=14$ T, while the intensities of the other trions components appear to be weakly affected by the magnetic field. These results highlight the fundamental differences between RC and PL investigations: effects photo-doping and of non-trivial carrier relaxation processes within the different energy levels can dominate the PL response.

The $\sigma^-$ polarized singlet trion appears as the only trion component which amplitude increases with the magnetic field. We understand this increase as a combined effect of charge transfer into the lowest K$^{-}$ valley, and of efficient carrier relaxation also towards energy states in the K$^{-}$ lowest valley, prior to optical recombination. The probability of trion formation involving free carriers in the K$^{+}$ valley strongly decreases with increasing magnetic fields due to the valley Zeeman effect and to the fixed Fermi energy, as evidenced in the RC experiments. This concerns the T$\unit{_T^{\sigma -}}$ and the T$\unit{_S^{\sigma +}}$ complexes which intensities are expected to vanish. T$\unit{_T^{\sigma +}}$ can also efficiently relax into T$\unit{_S^{\sigma -}}$ by flipping spins of both the electron and the hole, conserving in total both spin and momentum. These observations point towards a strongly enhanced T$\unit{_S^{\sigma -}}$ emission, with respect to other trion complexes.

The degree of circular polarization of the singlet trion increases linearly with the magnetic field, with a rate of $\sim 12.5 \% \cdot T^{-1}$, faster than that observed in the RC experiment. This would imply a smaller value for the Fermi energy in the PL experiment than in the RC experiment. When $\mathscr{P}$ reaches 80$\%$ at $B\sim6$ T, the rate decreases significantly down to ~$2.1 \% \cdot T^{-1}$ and $\mathscr{P}$ reaches 90$\%$ at $B=14$ T. The existence of these two regimes observed for the magnetic field induced evolution of the degree of circular polarization extracted from PL measurements is intriguing, and reflects mechanisms that cannot be elucidated with an continuous-wave photoluminescence experiments. Even if the evolution of the intensities of both triplet trion components appears as reduced, the degree of circular polarization shows first an increase up to $\sim 30 \%$ close to $B=4$ T, and then it vanishes. This evolution cannot be understood within the frame of magnetic field dependent free carrier concentration in the two valleys. It should result from a subtile interplay between different scattering processes of both radiative and non-radiative types, the effect of photodoping due to the excitation laser, and finally, to free carrier redistribution among the two valleys, which our experiment does not allow to identify.

\section{Conclusions}
To conclude, we have investigated the magnetic field evolution of negative trion resonances in a WS$_2$ monolayer encapsulated in hBN by means of reflectance contrast and photoluminescence experiments. These two techniques bring surprisingly different results. The RC response can be understood in terms carrier redistribution between the two valleys due to the relative shift of bands with respect to the fixed Fermi energy. This leads to a strong polarization of the trion complexes when applying a magnetic field, and also results in modified values for the g-factors characterizing the singlet and triplet trion species. In PL, a similar increase of polarization is only observed for the singlet trion, probably due to the interplay between competing effects of exciton relaxation to lower energy states and of changes in the probability of formation of a given trion when tuning the free carrier concentration in the two valleys. We extracted g-factors of both trion complexes and estimated electrons’ concentration. These findings show that magnetic fields enable to control the valley polarization of the free electrons in the conduction band, and highlight the complexity of the different mechanisms involved in the PL response.

\section*{Acknowledgements}
The work has been supported by the EU Graphene Flagship project, the ATOMOPTO project (TEAM programme of the Foundation for Polish Science, co-financed by the EU within the ERD-Fund), the National Science Centre, Poland (grants no. 2017/27/B/ST3/00205, 2017/27/N/ST3/01612, and 2018/31/B/ST3/02111), the Nano fab facility of the Institut N\'eel, CNRS UGA, and the LNCMI-CNRS, a member of the European Magnetic Field Laboratory (EMFL). P. K.  kindly acknowledges the National Science Centre, Poland (grant no. 2016/23/G/ST3/04114) for financial support for his PhD. M. B. acknowledges the financial support of the Ministry of Education, Youth and Sports of the Czech Republic under the project CEITEC 2020 (Grant No. LQ1601). K.W. and T.T acknowledge support from the Elemental Strategy Initiative conducted by the MEXT, Japan, Grant Number JPMXP0112101001,  JSPS KAKENHI Grant Numbers JP20H00354 and the CREST(JPMJCR15F3), JST.

\bibliographystyle{apsrev4-1}
\bibliography{Magneto_optics_trions_new}
	
\end{document}